\documentclass[%
 reprint,
 amsmath,amssymb,
 aps,
]{revtex4-2}

\usepackage{bbm}
\usepackage[utf8]{inputenc}
\usepackage{amssymb}
\usepackage{graphicx}
\usepackage{dcolumn}
\usepackage{bm}
\usepackage[table]{xcolor}
\usepackage{colortbl}
\usepackage[normalem]{ulem}

\begin{document}

\preprint{APS/123-QED}

\title{Graph-Based Feature Engineering to \\ Predict the Dynamical Properties of Condensed Matter}%

\author{An Wang}
\affiliation{Department of Chemistry, University of Warwick, Coventry CV4 7AL, United Kingdom}

\author{Gabriele C. Sosso}
\affiliation{Department of Chemistry, University of Warwick, Coventry CV4 7AL, United Kingdom}%
\email{G.Sosso@warwick.ac.uk}

\date{\today}

\begin{abstract}
We present a graph theory-based method to characterise flow defects and structural shifts in condensed matter. We explore the connection between dynamical properties, particularly the recently introduced concept of ``softness'', and graph-based features such as centrality and clustering coefficients. These topological features outperform conventional features based on Euclidean metric in predicting particles mobility and allow to correctly identify phase transitions as well. These results provide a new set of computational tools to investigate the dynamical properties of condensed matter systems.
\end{abstract}

\maketitle

Machine learning (ML) models can, in principle, be used to study the dynamical properties of condensed matter, enabling the identification of particles that play a critical role in macroscopic processes such as flow and fracture~\cite{alsayed2005premelting,taylor1934mechanism}. This task involves fitting a composite function $\bold{y} = f(h(\bold{x}))$, where $\bold{x}$ and $\bold{y}$ are the input samples and target property, respectively. Feature engineering via $\bold{X}=h(\bold{x})$ generates a feature matrix $\bold{X}$, and $\bold{y}=f(\bold{X})$ applies a ML model for prediction or classification. Feature engineering $h$ identifies key attributes affecting system behavior, while the model $f$ leverages these attributes to make predictions about the dynamical properties of specific particles - as well as about the dynamics of the system as a whole. Models that can capture the dynamics of the system via simple metrics, such as the concept of particles softness discussed below, represent an especially intriguing option - as they enhance the interpretability of the model whilst capturing the fundamental physics of the system.

One such model has been put forward by Cubuk \textit{et al.}~\cite{cubuk2015identifying,schoenholz2016structural}, who employed the symmetry functions introduced by Behler and Parrinello~\cite{behler2007generalized}:

\begin{enumerate}
    \item Radial structural function ($G$, ``Type A'' features hereafter)
    \begin{equation}
    G_I (i;r,\delta) = \sum_{j\in I} \text{e}^{-\frac{1}{2 \delta^2} (r-R_{ij})^2}
    \label{equ:g_func}
    \end{equation}
    
    \item Angular structural function ($\Psi$, ``Type B'' features hereafter)
    \begin{equation}
    \Psi_{IJ} (i; \xi,\lambda,\zeta) = \sum_{j \in I} \sum_{k \in J} \text{e}^{-\frac{R_{ij}^2+R_{jk}^2+R_{ik}^2}{\xi^2}} (1+\lambda \text{cos} \theta_{ijk})^{\zeta}
    \label{equ:psi_func}
    \end{equation}  
\end{enumerate}

\noindent to encode the local environment of the particles as $h$. Here, $I$ and $J$ are the chemical species being analyzed, $R_{ij}$ is the Euclidean distance between particles $i$ and $j$, $r$ the detection radius, $\delta$ is a measure of resolution, $\theta_{ijk}$ is the angle between vectors $\overrightarrow{\bold{R}}_{ij}$, $\overrightarrow{\bold{R}}_{ik}$, $\lambda = \pm1$ allows to focus on smaller or larger bond angles, and $\zeta$ sets the angular resolution. Interaction among particles $i$, $j$, $k$ is deemed significant if $R_{ij}^2+R_{jk}^2+R_{ik}^2<\xi^2$. Varying $I$, $J$, $r$, $\delta$, $\xi$, $\lambda$, and $\zeta$ yields a diverse portfolio of $G$ and $\Psi$ functions for all particles in $\bold{x}$, thus enabling the construction of the feature matrix $\bold{X}$.

According to Schoenholz \textit{et al.}~\cite{schoenholz2016structural}, we can quantify the extent to which the position of the $i$-th particle in a given system changes within a certain timescale $t_R$ via $p_{\text{hop},i}(t)$~\cite{schoenholz2016structural,candelier2010spatiotemporal,smessaert2013distribution}, which is defined as:

\begin{equation}
    p_{\text{hop},i}(t) = \sqrt{\langle (\bold{r}_i-\langle \bold{r}_i \rangle_B)^2 \rangle_A \langle (\bold{r}_i-\langle \bold{r}_i \rangle_A)^2 \rangle_B}
    \label{equ:p_hop_t}
\end{equation}

\noindent within the interval of $A \cup B$, where $A=[t-t_{R/2}, t]$ and $B=[t, t+t_{R/2}]$. Importantly, $p_{\text{hop},i}(t)$ provides a measure of the so-called ``softness'' of a particle $i$ at time $t$. In fact, we define a particle as either ``soft'' - if $p_{\text{hop},i}(t)$ exceeds a certain threshold $p_c$ within the $[t-t_{R/2},t+t_{R/2}]$ interval - or ``hard'' otherwise~\cite{schoenholz2016structural}.

Thus, for the purposes of building a ML model to characterise the dynamics of the system, we can label each particle as ``rearrangement-prone'' (or soft, $y_i=1$) or ``stable'' (or hard, $y_i=0$), where $y_i$ is the $i$-th element of $\bold{x}$. Note that adjusting the threshold setting $p_c$ can impact the outcomes of the trained model, for instance by causing a logarithmic shift in the energy scale which impacts the accuracy of the ML results. However, the choice of $p_c$ does not alter the fundamental insights into the dynamic properties of each particle, nor does it affect the discovery of basic physical laws~\cite{schoenholz2016structural}.

In the work of Cubuk \textit{et al}~\cite{cubuk2015identifying,schoenholz2016structural}, linear Support Vector Machines (SVM)~\cite{cortes1995support} were used to build a ML model connecting the local environment of each particle to its dynamical properties. In this context, the SVM seeks to fit the best hyperplane in input space, using a decision function defined as:

\begin{equation}
    f(\bold{X}) = \bold{X} \cdot \bold{w} + b,
    \label{equ:svm_linear}
\end{equation}

\noindent where $\bold{w}$ represents the weights and $b$ is the bias term. The softness of particle $i$ is defined as the directed distance between the $i$-th sample $\bold{X}_{(i)}$ and the decision plane $\bold{X} \cdot \bold{w} + b - \bold{y} = \bold{0}$: 

\begin{equation}
    \mathop{Softness}(i) = \frac{\bold{X}_{(i)} \cdot \bold{w}+b}{||\bold{w}||_2}
    \label{equ:softness}
\end{equation}

Hence, the changes in the physical properties of the particles are described through the evolution of a scalar field, i.e, the softness. In other words, the concept of softness quantifies the extent to which interactions drive changes in the positions of the particles, thus offering a unique angle to investigate the dynamics of disordered systems~\cite{cubuk2015identifying}. Note that a high number of nearest neighbors within the local atomic environments (identified via the intensity of the first peak of the radial distribution) leads to low values of particles softness~\cite{schoenholz2016structural}. This explains the predictive power of the symmetry function $G$, which can be considered as a discrete version of the radial distribution function, i.e., $g(r) = \frac{1}{4 \pi r^2} \langle \sum_{j \in I} \delta_D (R_{ij}-r) \rangle_i = \lim_{\delta \rightarrow 0} \frac{\langle G_I (i;r,\delta) \rangle_i}{4 \pi r^2}$.

The correspondence between the $G$ function and $g(r)$, rooted in the Euclidean geometry of $\mathbb{R}^3$, highlights the sensitivity of this metric to the specific phase of matter and/or the thermodynamic conditions, such as temperature or pressure. Cubuk \textit{et al.}~\cite{cubuk2015identifying,schoenholz2016structural} noted that scaling the target variable according to the temperature can mitigate this issue to some extent. However, this workaround is not sufficient for the model to be transferable across, e.g., different densities, or when the system undergoes a phase transition~\cite{cubuk2015identifying}. Moreover, such adjustments alter the criteria for classifying particles as soft or hard.

Here, we leverage the topology of the system, as opposed to features from Euclidean geometry (such as $G$ and $\Psi$), to uncover fundamental, metric-independent properties of the system. In particular, we adopt graph theory to provide a new set of tools to predict the dynamical properties of condensed matter systems via ML models that are transferable across different thermodynamic conditions and can be even used to identify the onset of phase transitions. Our approach involves two steps: (1) graphs generation, and (2) topological feature engineering on these graphs.

Firstly, we utilise a modified Voronoi method~\cite{malins2013identification} (discussed in detail in the [joint PRE article]) to generate a graph from a given configuration of the system. Consider a set of particles $\mathcal{V} = \{ v_1, \cdots, v_{\mathcal{N}}\}$ with  coordinates $\overrightarrow{v}_i$ in $\mathbb{R}^3$, and $\overrightarrow{r} = \overrightarrow{v}_j - \overrightarrow{v}_i$. The algorithm to add edges to $\mathcal{E}$, which is initialized to an empty set~\cite{malins2013identification}, is:
\begin{enumerate}
    \item Loop over all $\mathcal{N}$ particles with index $i$.
    \item Identify all particles within a distance $r_c$ of $\overrightarrow{r}_i$, ensuring $r_c$ exceeds the maximum bond length in the network, and include these in set $\mathcal{S}_i$.
    \item Sort the particles in $\mathcal{S}_i$ by their ascending distance from the $i$-th particle.
    \item Loop over all $v_j \in \mathcal{S}_i$, i.e., particles in increasing distance from the $i$-th particle.
    \item For each $j$ loop over all $k > j$ in $\mathcal{S}_i$ and eliminate $k$-th particle from $\mathcal{S}_i$ if the following inequality is not satisfied:
    \begin{equation}
        \mathcal{A} > \frac{|\overrightarrow{r}_{ik}|^2}{|\overrightarrow{r}_{ij}|^2+|\overrightarrow{r}_{jk}|^2}
        \label{eq_str}
    \end{equation}
    Add an edge between $i$ and each node in $\mathcal{S}_i$ to $\mathcal{E}$.
\end{enumerate}

Then, the inner function is a mapping of $h:(\mathcal{V},\mathcal{E})^m \rightarrow \mathbb{R}^n$, with $m$ indicating the count of distinct $\mathcal{A}$ values chosen, and $n$ the column count of feature matrix $\bold{X}$. This introduces two graph-based structural function types, each identifying ten predictors per $\mathcal{A}$.

The first type of features (``Type C'' features hereafter) consists of the heterogeneous descriptors composed of nine node centralities and the clustering coefficient, which, by incorporating a variety of different features, allows us to analyze data from multiple dimensions. Node centrality~\cite{rodrigues2019network,landherr2010critical}, a key concept in network science~\cite{barabasi2013network,albert2002statistical}, denotes the significance of a node within the network~\cite{borgatti2024analyzing}, and different centrality highlights varied aspects of this significance. The clustering coefficient~\cite{watts1998collective}, distinct from node centralities, underscores the local connectivity among the neighbors of a node. For node $i$, we incorporate: (1) Degree Centrality~\cite{freeman2002centrality}, i.e., $C_d(i)$; (2) H-index Centrality~\cite{hirsch2005index}, i.e., $H_i(i)$ is $h_i$; (3) Closeness Centrality~\cite{bavelas1950communication,freeman1977influence,wasserman1994social}, i.e., $C_C(i)$; (4) Betweenness Centrality~\cite{bavelas1948mathematical,freeman1977set}, i.e., $C_B(i)$; (5) Eigenvector Centrality~\cite{bonacich1972factoring,bonacich2007some}, i.e., $C_E(i)$; (6) K-shell Centrality~\cite{carmi2007model,kitsak2010identification}, i.e., $C_{KS}(i)$; (7) Clustering Coeffcient~\cite{watts1998collective}, i.e., $C_{CC}(i)$; (8) Subgraph Centrality~\cite{estrada2005subgraph}, i.e., $C_{SG}(i)$; (9) Harmonic Centrality~\cite{latora2007measure}, i.e., $C_H(i)$; (10) LocalRank Centrality~\cite{chen2012identifying}, i.e., $C_{LR}(i)$. The details relative to all these metrics are provided in the [joint PRE article]. Note that Type C features are entirely independent from Euclidean metric.

The second type of features (``Type D'' features hereafter) consists of the angular functions defined on the hierarchical structures of a node. The angular function of the $i$-th point about its $\mathcal{L}$-th neighbors $\mathcal{K}_{\mathcal{L}}$ under the strictness $\mathcal{A}$ can be defined as:

\begin{equation}
    \mathcal{F}_{\mathcal{A}}(i, \mathcal{L}) = \sum_{j \neq i, v_j \in \mathcal{K}_{\mathcal{L}}} \sum_{k \neq j \neq i, v_k \in \mathcal{K}_{\mathcal{L}}} (1+\text{cos}(\theta_{ijk})) \text{e}^{-\mathcal{L}}
    \label{equ:ang_on_graph}
\end{equation}

where $\theta_{ijk}$ means the angle between vectors $\overrightarrow{r}_{ij}$ and the $\overrightarrow{v_i}$ and $\overrightarrow{r}_{ij} = \overrightarrow{v}_j - \overrightarrow{v}_i$ comes from the metrics in original Euclidean space $\mathbb{R}^3$. Ten Type D descriptors can be obtained by iterating $\mathcal{L} = 1,2,\cdots,10$. Thus, $\mathcal{A}$ from Type C and D may be the same or vary within their combination. Note that Type D features do depend to an extent on Euclidean metric.

\begin{figure}[htbp]
\includegraphics[width=0.45\textwidth]{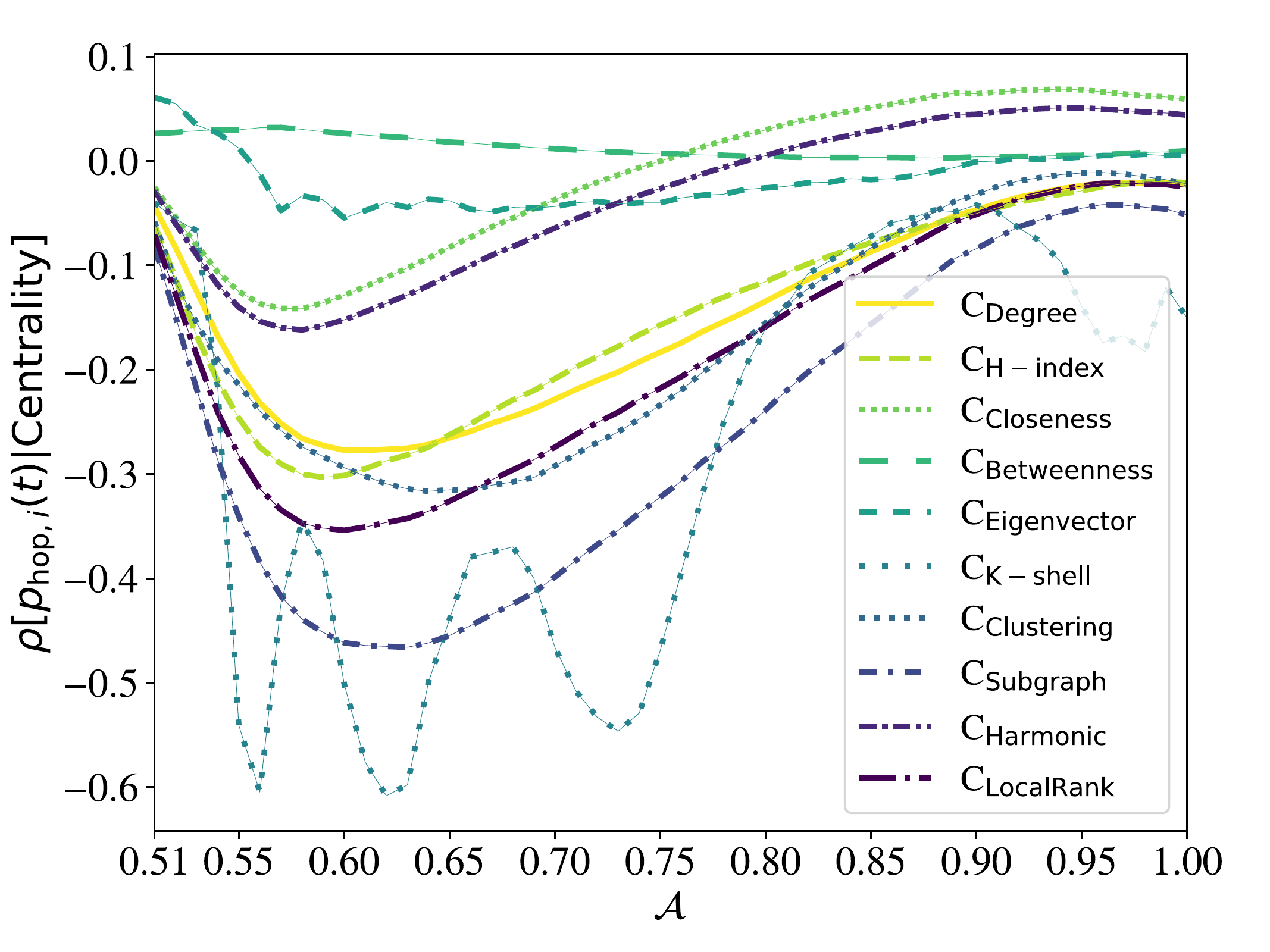}
\caption{The Spearman's rank correlation coefficient~\cite{spearman1987proof} between $p_{\text{hop,i}(t)}$ and Type C descriptors, i.e., the centralities and clustering coefficient under different $\mathcal{A}$, which is a dimensionless parameter to control the strictness of determining quadrilateral rings.}
\label{fig:p}
\end{figure}

In order to test the predictive capabilities of these graph-based features (Type C and D) compared to symmetry-based one (Type A and B), we focus on the potential crystallization of the (supercooled) Lennard-Jones (LJ) liquid. In particular, we run molecular dynamics (MD) simulations of a LJ system containing 864 particles and generate twenty trajectories at different temperatures. The details of the computational protocol can be found in the [jointed PRE article]. These results can be split into two groups: $\text{Traj}(T^{(1)}_i)=\text{Traj}(T_{2i-1})$ and $\text{Traj}(T^{(2)}_i)=\text{Traj}(T_{2i})$, $\forall i \in [1,10] \cap \mathbb{N}$. For each group, we collect all $864$ particle environments at $t_j = 100 \cdot j$-th steps ($j \in [1,9] \cap \mathbb{N}$) from their respective 10 trajectories to form two datasets: Set 1 (odd indexes), which we use for training the ML models, and Set 2 (even indexes) which we used as the test set instead. 

Intuitively, most particles in solids are expected to be hard, while most particles in liquids should be soft. Thus, we have classified a particle $i$ as soft (positive class, in the context of our ML models) if $p_{\text{hop},i}(t)$ (see Eq.~\ref{equ:p_hop_t}) exceeds $p_c = 0.1$ within the interval of $[t-20, t+20]$; otherwise, it's considered hard (negative class). These choices in terms of $p_c$ and $t_{R/2}$ are consistent with the results we have obtained in the [joint PRE article].

In disordered systems, the most crucial aspect of generating our graphs is the criterion utilised to decide whether or not to add an edge to the Voronoi graph. This criterion is the strictness $\mathcal{A}$ (see Eq.~\ref{eq_str}), which turns out to be much more impactful than the cutoff distance $r_c$ (as discussed in detail in the [joint PRE article]). Hence, here we fix $r_c=2.5$ and explore the impact of $\mathcal{A}$ (within the $(0.5,1]$ range).
As illustrated in Fig.~\ref{fig:p}, we find that both the centralities and clustering coefficients of particles correlate (to different extent) with $p_{\text{hop},i}(t)$. This is key, as it suggests that our graph-based metrics contain information that relates to the mobility of each particle. 

Broadly speaking, the greater correlations are observed for $\mathcal{A}$ between $\sim$0.55 and  $\sim$0.75. Eigenvector and Betweenness Centralities both show a very weak correlation with $p_{\text{hop},i}(t)$ across the whole range of $\mathcal{A}$ we have considered. Closeness and Harmonic Centralities exhibit a weak correlation with $p_{\text{hop},i}(t)$, especially for $\mathcal{A} \in [0.55,0.6]$. Degree, H-index, LocalRank Centralities, and Clustering Coefficient demonstrate a moderate correlation with $p_{\text{hop},i}(t)$ for $\mathcal{A} \in [0.55,0.7]$. Strong correlations with $p_{\text{hop},i}(t)$ are seen with K-shell and Subgraph Centralities for $\mathcal{A} \in [0.55,0.75]$. Notably, Subgraph Centrality around $\mathcal{A} = 0.6$ and K-shell Centrality showing periodic oscillations in the correlations with $p_{\text{hop},i}(t)$. Interestingly, we note that the K-shell centrality is the only metric among all 10 Type C descriptors that contains information about the core-periphery structure of the network. The periodic oscillations in the correlation between K-shell centrality and $p_{\text{hop},i}(t)$ occur because, as parameter $\mathcal{A}$ changes, the critical conditions determining whether there is an edge between nodes change. This, in turn, causes edges to be periodically added or removed, which periodically affects the core-periphery level of nodes, particularly those that are deemed to be significant. Based on these results, we have utilised different values of $\mathcal{A}$ to craft a diverse portfolio of graph-based features (as discussed in detail in [the joint PRE article]).

To build our ML model, aimed at predicting whether a given particle is soft or hard, we have randomly selected $15,000$ class-balanced samples from Set $1$ for the development set and $30,000$ from Set $2$ for the test set. In addition to the linear SVM~\cite{cortes1995support}, we have also explored the performance of nonlinear ML models, particularly SVMs with RBF kernels~\cite{buhmann2000radial,scholkopf2002learning}, ensemble learning (by combining multiple decision algorithms), and neural networks~\cite{mcculloch1943logical,rosenblatt1958perceptron,rumelhart1986learning}. We employed a 10-fold cross-validation on Set $1$ and a Bayesian optimizer~\cite{snoek2012practical} to fine-tune the hyperparameters of the relevant models. A more in-depth discussion relative to the ML models is provided in [the joint PRE article]. 
We have chosen the commonly used Matthews Correlation Coefficient (MCC)~\cite{matthews1975comparison} as the main metric to assess the accuracy of our predictions.

\begin{table}[htbp]
\footnotesize
\renewcommand{\arraystretch}{1.5}
\centering
\caption{Comparison between symmetry function-based descriptors (Type A and B) and graph-based descriptors (Type C and D). ``Vars" refer to the number of predictors, whilst ``ConfMat" refer to the confusion matrix, reported as percentages.}
\label{tab:b_1}
\begin{tabular}{c|c|c|c|c|c}
\hline
\multicolumn{6}{c}{Linear SVM} \\
\hline
Type & Parameters & Vars & Accuracy & MCC & ConfMat \\ \hline
\rowcolor{yellow!20}
A & $r\in[1,7.9]$, $\delta=0.1$ & $80$ & $71.7\%$ & $0.434$ & $\left(\begin{smallmatrix} 68.8\% & 31.2\% \\ 25.5\% & 74.5\% \end{smallmatrix}\right)$ \\
C & $\mathcal{A}=0.55$ & $10$ & $76.5\%$ & $0.530$ & $\left(\begin{smallmatrix} 75.2\% & 24.8\% \\ 22.1\% & 77.9\%\end{smallmatrix}\right)$ \\
\rowcolor{gray!20}
C & $\mathcal{A}=0.55,0.6,\cdots,0.7$ & $40$ & $83.2\%$ & $0.664$ & $\left(\begin{smallmatrix} 75.2\% & 24.8\% \\ 19.2\% & 80.8\% \end{smallmatrix}\right)$ \\
\hline
\rowcolor{yellow!20}
A+B & 80 A+20 diverse B & $100$ & $76.8\%$ & $0.535$ & $\left(\begin{smallmatrix} 77.4\% & 22.6\% \\ 23.8\% & 76.2\% \end{smallmatrix}\right)$ \\
C+D & $0.55$ for C, $0.6$ for D & $20$ & $82.6\%$ & $0.652$ & $\left(\begin{smallmatrix} 83.6\% & 16.4\% \\ 18.5\% & 81.5\% \end{smallmatrix}\right)$ \\
\rowcolor{gray!20}
C+D & $\mathcal{A}=0.55,0.6,\cdots,0.75$ & $100$ & $87.5\%$ & $0.750$ & $\left(\begin{smallmatrix} 89.0\% & 11.0\% \\ 14.1\% & 85.9\% \end{smallmatrix}\right)$  \\
\hline
\multicolumn{6}{c}{Selected models with the best performance} \\
\hline
Type & Model & Vars & Accuracy & MCC & ConfMat\\ \hline
\rowcolor{yellow!20}
A & SVM (RBF kernel) & $80$ & $82.8\%$ & $0.656$ & $\left(\begin{smallmatrix} 82.0\% & 18.0\% \\ 16.4\% & 83.6\% \end{smallmatrix}\right)$ \\
C & Ensemble Learning & $10$ & $85.7\%$ & $0.715$ & $\left(\begin{smallmatrix} 89.4\% & 10.6\% \\ 18.1\% & 81.9\% \end{smallmatrix}\right)$ \\
\rowcolor{gray!20}
C & Ensemble Learning & $40$ & $90.8\%$ & $0.817$ & $\left(\begin{smallmatrix} 92.6\% & 7.4\% \\ 10.9\% & 89.1\% \end{smallmatrix}\right)$ \\
\hline
\rowcolor{yellow!20}
A+B & SVM (RBF kernel) & $100$ & $84.8\%$ & $0.696$ & $\left(\begin{smallmatrix} 83.9\% & 16.1\% \\ 14.3\% & 85.7\% \end{smallmatrix}\right)$ \\
C+D & Ensemble Learning & $20$ & $86.6\%$ & $0.733$ & $\left(\begin{smallmatrix} 90.1\% & 9.9\% \\ 17.0\% & 83.0\% \end{smallmatrix}\right)$ \\
\rowcolor{gray!20}
C+D & Ensemble Learning & $100$ &$91.2\%$ & $0.824$ & $\left(\begin{smallmatrix} 92.8\% & 7.2\% \\ 10.4\% & 89.6\% \end{smallmatrix}\right)$ \\
\hline
\end{tabular}
\end{table}

\begin{figure*}[htbp!]
	\includegraphics[width=1.0\textwidth]{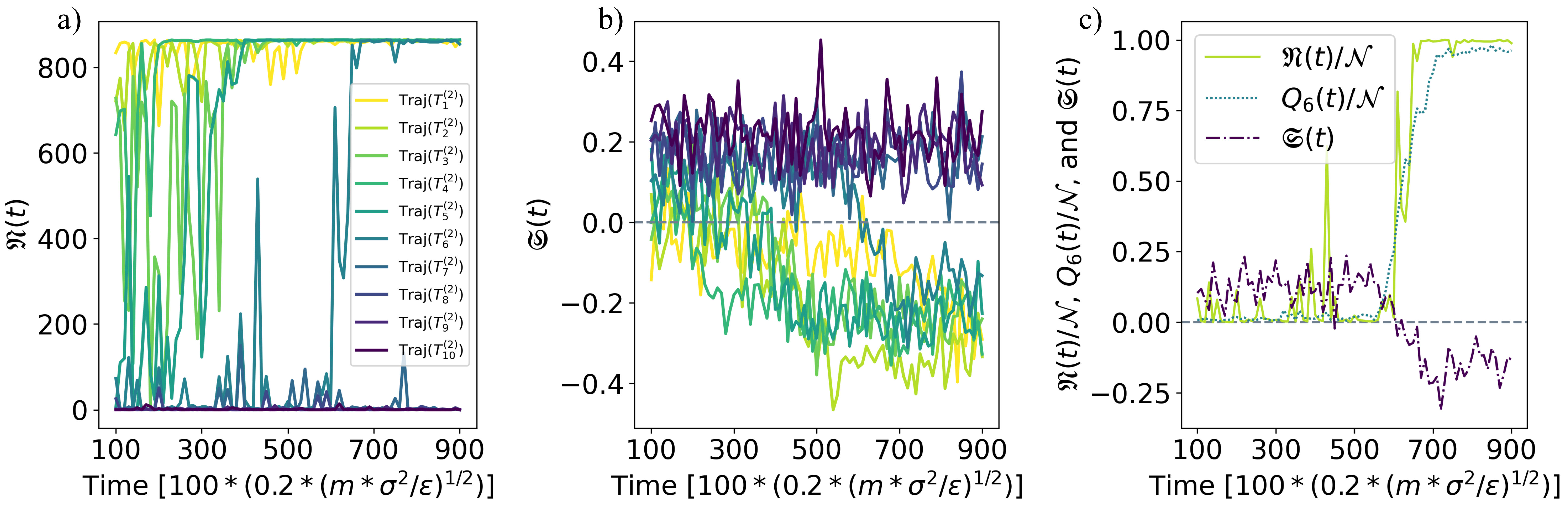}
	\caption{The a) hard particle counter $\mathfrak{N}(t)$ and b) average softness $\mathfrak{S}(t)$ on all trajectories from $\{\text{Traj} (T^{(2)}_p) | p\in[1,10] \cap \mathbb{N}\}$. c) Normalized $\mathfrak{N}(t)/\mathcal{N}$, normalized $Q_6(t)/\mathcal{N}$ and $\mathfrak{S}(t)$ for the selected trajectory of $\text{Traj}(T^{(2)}_6)$, where $\mathcal{N} = 864$. }
	\label{fig:softness} 
\end{figure*}

Our aim is to compare the performance of the symmetry functions-based descriptors (Type A and B, see Eqs.~\ref{equ:g_func} and ~\ref{equ:psi_func}) utilised in Ref.~\cite{cubuk2015identifying,schoenholz2016structural} with that of the graph-based descriptors (Type C and D) introduced in this work.
The results are summarised in Table~\ref{tab:b_1}, which include the results of linear SVMs, and the results of the best-performing non-linear models. The full set of results is available in [the joint PRE article]. We find that our graph-based descriptors consistently outperform the symmetry functions-based descriptors. As an example, using 80 Type A features results in a MCC of 0.434 (or 0.656 if utilising non linear SVMs), to be compared against a MCC score of 0.530 (or 0.715 if utilising non linear SVMs) obtained by using only 10 Type C descriptors. We note that optimizing a single $\mathcal{A}$ seems to be leading to best numerical accuracy, with optimal values being $0.55$ for Type C and $0.6$ for Type D (values that are consistent with the results reported in Fig.~\ref{fig:p}). Crucially, we note that whilst feature engineering can enhance the ability to capture complex relationships, reduce over fitting, and boost generalization, it cannot alter the inherent dependencies present in non-i.i.d. (i.e, Non-Independent and Non-Identically Distributed) data. 

Next, we apply our framework to the investigation of the crystallisation process of the supercooled liquid phase of the LJ system. To this end, we selected our top-performing model, i.e., the ensemble learning with $100$ ``C+D" predictors (see Table~\ref{tab:b_1}), to classify particles as either soft or hard, which allows us to define the following hard-particle counter:

\begin{equation}
    \mathfrak{N}(t) = \sum_{i=1}^{\mathcal{N}} \mathbbm{1}_{\text{particle } i \text{ is hard}}(t).
    \label{equ:counter}
\end{equation}

To define the system's average softness $\mathfrak{S}$ instead:

\begin{equation}
    \mathfrak{S}(t) = \frac{1}{\mathcal{N}} \sum_{i=1}^{\mathcal{N}} \frac{\bold{X}_{(i)}(t) \cdot \bold{w} + b}{||\bold{w}||_2},
    \label{equ:overall_softness}
\end{equation}

\noindent we have chosen the best performing linear SVM (to be consistent with the work of Cubuk \textit{et al.}~\cite{cubuk2015identifying,schoenholz2016structural}), which uses $100$ ``C+D" predictors (see Table~\ref{tab:b_1}). Here, $\mathcal{N}=864$ is the system size, $\mathbbm{1}_{\text{condition}}$ is the indicator function, $\bold{w}$ is the trained weights of linear SVM, and $\bold{X}_{(i)}(t)$ is the feature matrix of particle $i$ at time $t$. We applied $\mathfrak{N}(t)$ and $\mathfrak{S}(t)$ to the MD trajectories corresponding to Set 2, with sampling time $t_p = 100 + 10p$, where $p$ iterates over the interval of $[0, 80] \cap \mathbb{N}$.

The significance of $\mathfrak{N}(t)$ and $\mathfrak{S}(t)$ lies in their ability to link the local environments of particles with the global properties by taking the arithmetic average of the states of individual particles. Both $\mathfrak{N}(t)$ and $\mathfrak{S}(t)$ are extensive quantities, with very intuitive meanings. Thus, the evolution of the overall properties of the system can be effectively described by a new combinatorial counter, i.e., $\mathfrak{N}(t)$ or a new scalar field, i.e., $\mathfrak{S}(t)$, which result from this averaging process. Within this framework, the overall fluidity of the system is thus defined as the sum of the fluid properties of each particle, and $\mathfrak{N}(t)$ and $\mathfrak{S}(t)$ effectively serve as order parameters.

We find that, within the timescales probed via our MD simulations, the system crystallises for $T^{(1)}_j$, with $j\leq6$ - whilst it remains in the liquid state for $T^{(1)}_k$, with $k \geq 7$.
Both $\mathfrak{N}(t)$ and $\mathfrak{S}(t)$ effectively distinguish between systems that ultimately solidify or liquefy, as shown in Fig.~\ref{fig:softness}a) and Fig.~\ref{fig:softness}b), respectively. For systems solidifying, $\mathfrak{N}(t)$ stabilizes near $\mathcal{N}$ with $\mathfrak{S}(t) < 0$; for those liquefying, $\mathfrak{N}(t)$ nears $0$ as $\mathfrak{S}(t) > 0$. For trajectories encapsulating the full crystallization process from a supercooled liquid (such as $\text{Traj}(T^{(2)}_6)$, reported in Fig.~\ref{fig:softness}c), both $\mathfrak{N}(t)$ and $\mathfrak{S}(t)$ are in excellent agreement with the results obtained via the ten Wolde order parameter $Q_6(t)$. Note that, $\mathfrak{N}(t)$ and $\mathfrak{S}(t)$ assess crystallization in supercooled liquids by monitoring fluidity changes, while $Q_6$ evaluates it by detecting the formation of long-range order (LRO). In Classical Nucleation Theory (CNT)~\cite{karthika2016review,smeets2017classical}, crystallization begins with a stable nucleus forming. When it exceeds a critical size, it tends to grow rather than shrink, allowing the nucleus to grow. Thus, in systems crystallising from a supercooled liquid, $Q_6(t)$ should indeed align with $\mathcal{N}$. The shift of $\mathfrak{S}(t)$ from positive (liquid/soft) to negative (crystalline/hard) values at the onset of the crystallization process is less stark if compared to the change in $\mathfrak{N}(t)$ (as $\mathfrak{N}(t)$ is a scalar field as opposed to a counter) but equally consistent with the evolution of $Q_6(t)$.

In conclusion, we have presented a graph-based method for analyzing the dynamical properties of condensed matter, with emphasis on the recently introduced concept of ``softness''. We have constructed a set of topological, graph-based descriptors that outperform traditional symmetry functions in predicting the mobility of a given particle. Our approach offers an intriguing alternative to the state-of-the-art descriptors used in the context of ML for condensed matter, which - in most cases - heavily rely on Euclidean metric as opposed to topological features. In addition, our graph-based framework accurately characterise the crystallization of the liquid phase across a variety of thermodynamic conditions. Importantly, this set of theoretical and computational tools can in principle be used to characterise any other property of a condensed matter system, which paves the way for future applications aimed at harnessing graph-based features in conjunction with traditional metrics.

We gratefully acknowledge the use of the high-performance computing (HPC) facilities provided by the Scientific Computing Research Technology Platform (SCRTP) at the University of Warwick.

\newpage


\begin{thebibliography}{39}%
\makeatletter
\providecommand \@ifxundefined [1]{%
 \@ifx{#1\undefined}
}%
\providecommand \@ifnum [1]{%
 \ifnum #1\expandafter \@firstoftwo
 \else \expandafter \@secondoftwo
 \fi
}%
\providecommand \@ifx [1]{%
 \ifx #1\expandafter \@firstoftwo
 \else \expandafter \@secondoftwo
 \fi
}%
\providecommand \natexlab [1]{#1}%
\providecommand \enquote  [1]{``#1''}%
\providecommand \bibnamefont  [1]{#1}%
\providecommand \bibfnamefont [1]{#1}%
\providecommand \citenamefont [1]{#1}%
\providecommand \href@noop [0]{\@secondoftwo}%
\providecommand \href [0]{\begingroup \@sanitize@url \@href}%
\providecommand \@href[1]{\@@startlink{#1}\@@href}%
\providecommand \@@href[1]{\endgroup#1\@@endlink}%
\providecommand \@sanitize@url [0]{\catcode `\\12\catcode `\$12\catcode `\&12\catcode `\#12\catcode `\^12\catcode `\_12\catcode `\%12\relax}%
\providecommand \@@startlink[1]{}%
\providecommand \@@endlink[0]{}%
\providecommand \url  [0]{\begingroup\@sanitize@url \@url }%
\providecommand \@url [1]{\endgroup\@href {#1}{\urlprefix }}%
\providecommand \urlprefix  [0]{URL }%
\providecommand \Eprint [0]{\href }%
\providecommand \doibase [0]{https://doi.org/}%
\providecommand \selectlanguage [0]{\@gobble}%
\providecommand \bibinfo  [0]{\@secondoftwo}%
\providecommand \bibfield  [0]{\@secondoftwo}%
\providecommand \translation [1]{[#1]}%
\providecommand \BibitemOpen [0]{}%
\providecommand \bibitemStop [0]{}%
\providecommand \bibitemNoStop [0]{.\EOS\space}%
\providecommand \EOS [0]{\spacefactor3000\relax}%
\providecommand \BibitemShut  [1]{\csname bibitem#1\endcsname}%
\let\auto@bib@innerbib\@empty
\bibitem [{\citenamefont {Alsayed}\ \emph {et~al.}(2005)\citenamefont {Alsayed}, \citenamefont {Islam}, \citenamefont {Zhang}, \citenamefont {Collings},\ and\ \citenamefont {Yodh}}]{alsayed2005premelting}%
  \BibitemOpen
  \bibfield  {author} {\bibinfo {author} {\bibfnamefont {A.~M.}\ \bibnamefont {Alsayed}}, \bibinfo {author} {\bibfnamefont {M.~F.}\ \bibnamefont {Islam}}, \bibinfo {author} {\bibfnamefont {J.}~\bibnamefont {Zhang}}, \bibinfo {author} {\bibfnamefont {P.~J.}\ \bibnamefont {Collings}},\ and\ \bibinfo {author} {\bibfnamefont {A.~G.}\ \bibnamefont {Yodh}},\ }\bibfield  {title} {\bibinfo {title} {Premelting at defects within bulk colloidal crystals},\ }\href@noop {} {\bibfield  {journal} {\bibinfo  {journal} {Science}\ }\textbf {\bibinfo {volume} {309}},\ \bibinfo {pages} {1207} (\bibinfo {year} {2005})}\BibitemShut {NoStop}%
\bibitem [{\citenamefont {Taylor}(1934)}]{taylor1934mechanism}%
  \BibitemOpen
  \bibfield  {author} {\bibinfo {author} {\bibfnamefont {G.~I.}\ \bibnamefont {Taylor}},\ }\bibfield  {title} {\bibinfo {title} {The mechanism of plastic deformation of crystals. part i.—theoretical},\ }\href@noop {} {\bibfield  {journal} {\bibinfo  {journal} {Proceedings of the Royal Society of London. Series A, Containing Papers of a Mathematical and Physical Character}\ }\textbf {\bibinfo {volume} {145}},\ \bibinfo {pages} {362} (\bibinfo {year} {1934})}\BibitemShut {NoStop}%
\bibitem [{\citenamefont {Cubuk}\ \emph {et~al.}(2015)\citenamefont {Cubuk}, \citenamefont {Schoenholz}, \citenamefont {Rieser}, \citenamefont {Malone}, \citenamefont {Rottler}, \citenamefont {Durian}, \citenamefont {Kaxiras},\ and\ \citenamefont {Liu}}]{cubuk2015identifying}%
  \BibitemOpen
  \bibfield  {author} {\bibinfo {author} {\bibfnamefont {E.~D.}\ \bibnamefont {Cubuk}}, \bibinfo {author} {\bibfnamefont {S.~S.}\ \bibnamefont {Schoenholz}}, \bibinfo {author} {\bibfnamefont {J.~M.}\ \bibnamefont {Rieser}}, \bibinfo {author} {\bibfnamefont {B.~D.}\ \bibnamefont {Malone}}, \bibinfo {author} {\bibfnamefont {J.}~\bibnamefont {Rottler}}, \bibinfo {author} {\bibfnamefont {D.~J.}\ \bibnamefont {Durian}}, \bibinfo {author} {\bibfnamefont {E.}~\bibnamefont {Kaxiras}},\ and\ \bibinfo {author} {\bibfnamefont {A.~J.}\ \bibnamefont {Liu}},\ }\bibfield  {title} {\bibinfo {title} {Identifying structural flow defects in disordered solids using machine-learning methods},\ }\href@noop {} {\bibfield  {journal} {\bibinfo  {journal} {Physical review letters}\ }\textbf {\bibinfo {volume} {114}},\ \bibinfo {pages} {108001} (\bibinfo {year} {2015})}\BibitemShut {NoStop}%
\bibitem [{\citenamefont {Schoenholz}\ \emph {et~al.}(2016)\citenamefont {Schoenholz}, \citenamefont {Cubuk}, \citenamefont {Sussman}, \citenamefont {Kaxiras},\ and\ \citenamefont {Liu}}]{schoenholz2016structural}%
  \BibitemOpen
  \bibfield  {author} {\bibinfo {author} {\bibfnamefont {S.~S.}\ \bibnamefont {Schoenholz}}, \bibinfo {author} {\bibfnamefont {E.~D.}\ \bibnamefont {Cubuk}}, \bibinfo {author} {\bibfnamefont {D.~M.}\ \bibnamefont {Sussman}}, \bibinfo {author} {\bibfnamefont {E.}~\bibnamefont {Kaxiras}},\ and\ \bibinfo {author} {\bibfnamefont {A.~J.}\ \bibnamefont {Liu}},\ }\bibfield  {title} {\bibinfo {title} {A structural approach to relaxation in glassy liquids},\ }\href@noop {} {\bibfield  {journal} {\bibinfo  {journal} {Nature Physics}\ }\textbf {\bibinfo {volume} {12}},\ \bibinfo {pages} {469} (\bibinfo {year} {2016})}\BibitemShut {NoStop}%
\bibitem [{\citenamefont {Behler}\ and\ \citenamefont {Parrinello}(2007)}]{behler2007generalized}%
  \BibitemOpen
  \bibfield  {author} {\bibinfo {author} {\bibfnamefont {J.}~\bibnamefont {Behler}}\ and\ \bibinfo {author} {\bibfnamefont {M.}~\bibnamefont {Parrinello}},\ }\bibfield  {title} {\bibinfo {title} {Generalized neural-network representation of high-dimensional potential-energy surfaces},\ }\href@noop {} {\bibfield  {journal} {\bibinfo  {journal} {Physical review letters}\ }\textbf {\bibinfo {volume} {98}},\ \bibinfo {pages} {146401} (\bibinfo {year} {2007})}\BibitemShut {NoStop}%
\bibitem [{\citenamefont {Candelier}\ \emph {et~al.}(2010)\citenamefont {Candelier}, \citenamefont {Widmer-Cooper}, \citenamefont {Kummerfeld}, \citenamefont {Dauchot}, \citenamefont {Biroli}, \citenamefont {Harrowell},\ and\ \citenamefont {Reichman}}]{candelier2010spatiotemporal}%
  \BibitemOpen
  \bibfield  {author} {\bibinfo {author} {\bibfnamefont {R.}~\bibnamefont {Candelier}}, \bibinfo {author} {\bibfnamefont {A.}~\bibnamefont {Widmer-Cooper}}, \bibinfo {author} {\bibfnamefont {J.~K.}\ \bibnamefont {Kummerfeld}}, \bibinfo {author} {\bibfnamefont {O.}~\bibnamefont {Dauchot}}, \bibinfo {author} {\bibfnamefont {G.}~\bibnamefont {Biroli}}, \bibinfo {author} {\bibfnamefont {P.}~\bibnamefont {Harrowell}},\ and\ \bibinfo {author} {\bibfnamefont {D.~R.}\ \bibnamefont {Reichman}},\ }\bibfield  {title} {\bibinfo {title} {Spatiotemporal hierarchy of relaxation events, dynamical heterogeneities, and structural reorganization in a supercooled liquid},\ }\href@noop {} {\bibfield  {journal} {\bibinfo  {journal} {Physical review letters}\ }\textbf {\bibinfo {volume} {105}},\ \bibinfo {pages} {135702} (\bibinfo {year} {2010})}\BibitemShut {NoStop}%
\bibitem [{\citenamefont {Smessaert}\ and\ \citenamefont {Rottler}(2013)}]{smessaert2013distribution}%
  \BibitemOpen
  \bibfield  {author} {\bibinfo {author} {\bibfnamefont {A.}~\bibnamefont {Smessaert}}\ and\ \bibinfo {author} {\bibfnamefont {J.}~\bibnamefont {Rottler}},\ }\bibfield  {title} {\bibinfo {title} {Distribution of local relaxation events in an aging three-dimensional glass: Spatiotemporal correlation and dynamical heterogeneity},\ }\href@noop {} {\bibfield  {journal} {\bibinfo  {journal} {Physical Review E}\ }\textbf {\bibinfo {volume} {88}},\ \bibinfo {pages} {022314} (\bibinfo {year} {2013})}\BibitemShut {NoStop}%
\bibitem [{\citenamefont {Cortes}\ and\ \citenamefont {Vapnik}(1995)}]{cortes1995support}%
  \BibitemOpen
  \bibfield  {author} {\bibinfo {author} {\bibfnamefont {C.}~\bibnamefont {Cortes}}\ and\ \bibinfo {author} {\bibfnamefont {V.}~\bibnamefont {Vapnik}},\ }\bibfield  {title} {\bibinfo {title} {Support-vector networks},\ }\href@noop {} {\bibfield  {journal} {\bibinfo  {journal} {Machine learning}\ }\textbf {\bibinfo {volume} {20}},\ \bibinfo {pages} {273} (\bibinfo {year} {1995})}\BibitemShut {NoStop}%
\bibitem [{\citenamefont {Malins}\ \emph {et~al.}(2013)\citenamefont {Malins}, \citenamefont {Williams}, \citenamefont {Eggers},\ and\ \citenamefont {Royall}}]{malins2013identification}%
  \BibitemOpen
  \bibfield  {author} {\bibinfo {author} {\bibfnamefont {A.}~\bibnamefont {Malins}}, \bibinfo {author} {\bibfnamefont {S.~R.}\ \bibnamefont {Williams}}, \bibinfo {author} {\bibfnamefont {J.}~\bibnamefont {Eggers}},\ and\ \bibinfo {author} {\bibfnamefont {C.~P.}\ \bibnamefont {Royall}},\ }\bibfield  {title} {\bibinfo {title} {Identification of structure in condensed matter with the topological cluster classification},\ }\href@noop {} {\bibfield  {journal} {\bibinfo  {journal} {The Journal of chemical physics}\ }\textbf {\bibinfo {volume} {139}} (\bibinfo {year} {2013})}\BibitemShut {NoStop}%
\bibitem [{\citenamefont {Rodrigues}(2019)}]{rodrigues2019network}%
  \BibitemOpen
  \bibfield  {author} {\bibinfo {author} {\bibfnamefont {F.~A.}\ \bibnamefont {Rodrigues}},\ }\bibfield  {title} {\bibinfo {title} {Network centrality: an introduction},\ }\href@noop {} {\bibfield  {journal} {\bibinfo  {journal} {A mathematical modeling approach from nonlinear dynamics to complex systems}\ ,\ \bibinfo {pages} {177}} (\bibinfo {year} {2019})}\BibitemShut {NoStop}%
\bibitem [{\citenamefont {Landherr}\ \emph {et~al.}(2010)\citenamefont {Landherr}, \citenamefont {Friedl},\ and\ \citenamefont {Heidemann}}]{landherr2010critical}%
  \BibitemOpen
  \bibfield  {author} {\bibinfo {author} {\bibfnamefont {A.}~\bibnamefont {Landherr}}, \bibinfo {author} {\bibfnamefont {B.}~\bibnamefont {Friedl}},\ and\ \bibinfo {author} {\bibfnamefont {J.}~\bibnamefont {Heidemann}},\ }\bibfield  {title} {\bibinfo {title} {A critical review of centrality measures in social networks},\ }\href@noop {} {\bibfield  {journal} {\bibinfo  {journal} {Wirtschaftsinformatik}\ }\textbf {\bibinfo {volume} {52}},\ \bibinfo {pages} {367} (\bibinfo {year} {2010})}\BibitemShut {NoStop}%
\bibitem [{\citenamefont {Barab{\'a}si}(2013)}]{barabasi2013network}%
  \BibitemOpen
  \bibfield  {author} {\bibinfo {author} {\bibfnamefont {A.-L.}\ \bibnamefont {Barab{\'a}si}},\ }\bibfield  {title} {\bibinfo {title} {Network science},\ }\href@noop {} {\bibfield  {journal} {\bibinfo  {journal} {Philosophical Transactions of the Royal Society A: Mathematical, Physical and Engineering Sciences}\ }\textbf {\bibinfo {volume} {371}},\ \bibinfo {pages} {20120375} (\bibinfo {year} {2013})}\BibitemShut {NoStop}%
\bibitem [{\citenamefont {Albert}\ and\ \citenamefont {Barab{\'a}si}(2002)}]{albert2002statistical}%
  \BibitemOpen
  \bibfield  {author} {\bibinfo {author} {\bibfnamefont {R.}~\bibnamefont {Albert}}\ and\ \bibinfo {author} {\bibfnamefont {A.-L.}\ \bibnamefont {Barab{\'a}si}},\ }\bibfield  {title} {\bibinfo {title} {Statistical mechanics of complex networks},\ }\href@noop {} {\bibfield  {journal} {\bibinfo  {journal} {Reviews of modern physics}\ }\textbf {\bibinfo {volume} {74}},\ \bibinfo {pages} {47} (\bibinfo {year} {2002})}\BibitemShut {NoStop}%
\bibitem [{\citenamefont {Borgatti}\ \emph {et~al.}(2024)\citenamefont {Borgatti}, \citenamefont {Everett}, \citenamefont {Johnson},\ and\ \citenamefont {Agneessens}}]{borgatti2024analyzing}%
  \BibitemOpen
  \bibfield  {author} {\bibinfo {author} {\bibfnamefont {S.~P.}\ \bibnamefont {Borgatti}}, \bibinfo {author} {\bibfnamefont {M.~G.}\ \bibnamefont {Everett}}, \bibinfo {author} {\bibfnamefont {J.~C.}\ \bibnamefont {Johnson}},\ and\ \bibinfo {author} {\bibfnamefont {F.}~\bibnamefont {Agneessens}},\ }\href@noop {} {\emph {\bibinfo {title} {Analyzing social networks}}}\ (\bibinfo  {publisher} {SAGE Publications Limited},\ \bibinfo {year} {2024})\BibitemShut {NoStop}%
\bibitem [{\citenamefont {Watts}\ and\ \citenamefont {Strogatz}(1998)}]{watts1998collective}%
  \BibitemOpen
  \bibfield  {author} {\bibinfo {author} {\bibfnamefont {D.~J.}\ \bibnamefont {Watts}}\ and\ \bibinfo {author} {\bibfnamefont {S.~H.}\ \bibnamefont {Strogatz}},\ }\bibfield  {title} {\bibinfo {title} {Collective dynamics of ‘small-world’networks},\ }\href@noop {} {\bibfield  {journal} {\bibinfo  {journal} {nature}\ }\textbf {\bibinfo {volume} {393}},\ \bibinfo {pages} {440} (\bibinfo {year} {1998})}\BibitemShut {NoStop}%
\bibitem [{\citenamefont {Freeman}\ \emph {et~al.}(2002)\citenamefont {Freeman} \emph {et~al.}}]{freeman2002centrality}%
  \BibitemOpen
  \bibfield  {author} {\bibinfo {author} {\bibfnamefont {L.~C.}\ \bibnamefont {Freeman}} \emph {et~al.},\ }\bibfield  {title} {\bibinfo {title} {Centrality in social networks: Conceptual clarification},\ }\href@noop {} {\bibfield  {journal} {\bibinfo  {journal} {Social network: critical concepts in sociology. Londres: Routledge}\ }\textbf {\bibinfo {volume} {1}},\ \bibinfo {pages} {238} (\bibinfo {year} {2002})}\BibitemShut {NoStop}%
\bibitem [{\citenamefont {Hirsch}(2005)}]{hirsch2005index}%
  \BibitemOpen
  \bibfield  {author} {\bibinfo {author} {\bibfnamefont {J.~E.}\ \bibnamefont {Hirsch}},\ }\bibfield  {title} {\bibinfo {title} {An index to quantify an individual's scientific research output},\ }\href@noop {} {\bibfield  {journal} {\bibinfo  {journal} {Proceedings of the National academy of Sciences}\ }\textbf {\bibinfo {volume} {102}},\ \bibinfo {pages} {16569} (\bibinfo {year} {2005})}\BibitemShut {NoStop}%
\bibitem [{\citenamefont {Bavelas}(1950)}]{bavelas1950communication}%
  \BibitemOpen
  \bibfield  {author} {\bibinfo {author} {\bibfnamefont {A.}~\bibnamefont {Bavelas}},\ }\bibfield  {title} {\bibinfo {title} {Communication patterns in task-oriented groups},\ }\href@noop {} {\bibfield  {journal} {\bibinfo  {journal} {The journal of the acoustical society of America}\ }\textbf {\bibinfo {volume} {22}},\ \bibinfo {pages} {725} (\bibinfo {year} {1950})}\BibitemShut {NoStop}%
\bibitem [{\citenamefont {Freeman}(1977{\natexlab{a}})}]{freeman1977influence}%
  \BibitemOpen
  \bibfield  {author} {\bibinfo {author} {\bibfnamefont {E.~W.}\ \bibnamefont {Freeman}},\ }\bibfield  {title} {\bibinfo {title} {Influence of personality attributes on abortion experiences},\ }\href@noop {} {\bibfield  {journal} {\bibinfo  {journal} {American Journal of Orthopsychiatry}\ }\textbf {\bibinfo {volume} {47}},\ \bibinfo {pages} {503} (\bibinfo {year} {1977}{\natexlab{a}})}\BibitemShut {NoStop}%
\bibitem [{\citenamefont {Wasserman}(1994)}]{wasserman1994social}%
  \BibitemOpen
  \bibfield  {author} {\bibinfo {author} {\bibfnamefont {S.}~\bibnamefont {Wasserman}},\ }\bibfield  {title} {\bibinfo {title} {Social network analysis: methods and applications},\ }\href@noop {} {\bibfield  {journal} {\bibinfo  {journal} {Cambridge University Press google schola}\ }\textbf {\bibinfo {volume} {2}},\ \bibinfo {pages} {131} (\bibinfo {year} {1994})}\BibitemShut {NoStop}%
\bibitem [{\citenamefont {Bavelas}(1948)}]{bavelas1948mathematical}%
  \BibitemOpen
  \bibfield  {author} {\bibinfo {author} {\bibfnamefont {A.}~\bibnamefont {Bavelas}},\ }\bibfield  {title} {\bibinfo {title} {A mathematical model for group structures},\ }\href@noop {} {\bibfield  {journal} {\bibinfo  {journal} {Human organization}\ }\textbf {\bibinfo {volume} {7}},\ \bibinfo {pages} {16} (\bibinfo {year} {1948})}\BibitemShut {NoStop}%
\bibitem [{\citenamefont {Freeman}(1977{\natexlab{b}})}]{freeman1977set}%
  \BibitemOpen
  \bibfield  {author} {\bibinfo {author} {\bibfnamefont {L.~C.}\ \bibnamefont {Freeman}},\ }\bibfield  {title} {\bibinfo {title} {A set of measures of centrality based on betweenness},\ }\href@noop {} {\bibfield  {journal} {\bibinfo  {journal} {Sociometry}\ ,\ \bibinfo {pages} {35}} (\bibinfo {year} {1977}{\natexlab{b}})}\BibitemShut {NoStop}%
\bibitem [{\citenamefont {Bonacich}(1972)}]{bonacich1972factoring}%
  \BibitemOpen
  \bibfield  {author} {\bibinfo {author} {\bibfnamefont {P.}~\bibnamefont {Bonacich}},\ }\bibfield  {title} {\bibinfo {title} {Factoring and weighting approaches to status scores and clique identification},\ }\href@noop {} {\bibfield  {journal} {\bibinfo  {journal} {Journal of mathematical sociology}\ }\textbf {\bibinfo {volume} {2}},\ \bibinfo {pages} {113} (\bibinfo {year} {1972})}\BibitemShut {NoStop}%
\bibitem [{\citenamefont {Bonacich}(2007)}]{bonacich2007some}%
  \BibitemOpen
  \bibfield  {author} {\bibinfo {author} {\bibfnamefont {P.}~\bibnamefont {Bonacich}},\ }\bibfield  {title} {\bibinfo {title} {Some unique properties of eigenvector centrality},\ }\href@noop {} {\bibfield  {journal} {\bibinfo  {journal} {Social networks}\ }\textbf {\bibinfo {volume} {29}},\ \bibinfo {pages} {555} (\bibinfo {year} {2007})}\BibitemShut {NoStop}%
\bibitem [{\citenamefont {Carmi}\ \emph {et~al.}(2007)\citenamefont {Carmi}, \citenamefont {Havlin}, \citenamefont {Kirkpatrick}, \citenamefont {Shavitt},\ and\ \citenamefont {Shir}}]{carmi2007model}%
  \BibitemOpen
  \bibfield  {author} {\bibinfo {author} {\bibfnamefont {S.}~\bibnamefont {Carmi}}, \bibinfo {author} {\bibfnamefont {S.}~\bibnamefont {Havlin}}, \bibinfo {author} {\bibfnamefont {S.}~\bibnamefont {Kirkpatrick}}, \bibinfo {author} {\bibfnamefont {Y.}~\bibnamefont {Shavitt}},\ and\ \bibinfo {author} {\bibfnamefont {E.}~\bibnamefont {Shir}},\ }\bibfield  {title} {\bibinfo {title} {A model of internet topology using k-shell decomposition},\ }\href@noop {} {\bibfield  {journal} {\bibinfo  {journal} {Proceedings of the National Academy of Sciences}\ }\textbf {\bibinfo {volume} {104}},\ \bibinfo {pages} {11150} (\bibinfo {year} {2007})}\BibitemShut {NoStop}%
\bibitem [{\citenamefont {Kitsak}\ \emph {et~al.}(2010)\citenamefont {Kitsak}, \citenamefont {Gallos}, \citenamefont {Havlin}, \citenamefont {Liljeros}, \citenamefont {Muchnik}, \citenamefont {Stanley},\ and\ \citenamefont {Makse}}]{kitsak2010identification}%
  \BibitemOpen
  \bibfield  {author} {\bibinfo {author} {\bibfnamefont {M.}~\bibnamefont {Kitsak}}, \bibinfo {author} {\bibfnamefont {L.~K.}\ \bibnamefont {Gallos}}, \bibinfo {author} {\bibfnamefont {S.}~\bibnamefont {Havlin}}, \bibinfo {author} {\bibfnamefont {F.}~\bibnamefont {Liljeros}}, \bibinfo {author} {\bibfnamefont {L.}~\bibnamefont {Muchnik}}, \bibinfo {author} {\bibfnamefont {H.~E.}\ \bibnamefont {Stanley}},\ and\ \bibinfo {author} {\bibfnamefont {H.~A.}\ \bibnamefont {Makse}},\ }\bibfield  {title} {\bibinfo {title} {Identification of influential spreaders in complex networks},\ }\href@noop {} {\bibfield  {journal} {\bibinfo  {journal} {Nature physics}\ }\textbf {\bibinfo {volume} {6}},\ \bibinfo {pages} {888} (\bibinfo {year} {2010})}\BibitemShut {NoStop}%
\bibitem [{\citenamefont {Estrada}\ and\ \citenamefont {Rodriguez-Velazquez}(2005)}]{estrada2005subgraph}%
  \BibitemOpen
  \bibfield  {author} {\bibinfo {author} {\bibfnamefont {E.}~\bibnamefont {Estrada}}\ and\ \bibinfo {author} {\bibfnamefont {J.~A.}\ \bibnamefont {Rodriguez-Velazquez}},\ }\bibfield  {title} {\bibinfo {title} {Subgraph centrality in complex networks},\ }\href@noop {} {\bibfield  {journal} {\bibinfo  {journal} {Physical Review E}\ }\textbf {\bibinfo {volume} {71}},\ \bibinfo {pages} {056103} (\bibinfo {year} {2005})}\BibitemShut {NoStop}%
\bibitem [{\citenamefont {Latora}\ and\ \citenamefont {Marchiori}(2007)}]{latora2007measure}%
  \BibitemOpen
  \bibfield  {author} {\bibinfo {author} {\bibfnamefont {V.}~\bibnamefont {Latora}}\ and\ \bibinfo {author} {\bibfnamefont {M.}~\bibnamefont {Marchiori}},\ }\bibfield  {title} {\bibinfo {title} {A measure of centrality based on network efficiency},\ }\href@noop {} {\bibfield  {journal} {\bibinfo  {journal} {New Journal of Physics}\ }\textbf {\bibinfo {volume} {9}},\ \bibinfo {pages} {188} (\bibinfo {year} {2007})}\BibitemShut {NoStop}%
\bibitem [{\citenamefont {Chen}\ \emph {et~al.}(2012)\citenamefont {Chen}, \citenamefont {L{\"u}}, \citenamefont {Shang}, \citenamefont {Zhang},\ and\ \citenamefont {Zhou}}]{chen2012identifying}%
  \BibitemOpen
  \bibfield  {author} {\bibinfo {author} {\bibfnamefont {D.}~\bibnamefont {Chen}}, \bibinfo {author} {\bibfnamefont {L.}~\bibnamefont {L{\"u}}}, \bibinfo {author} {\bibfnamefont {M.-S.}\ \bibnamefont {Shang}}, \bibinfo {author} {\bibfnamefont {Y.-C.}\ \bibnamefont {Zhang}},\ and\ \bibinfo {author} {\bibfnamefont {T.}~\bibnamefont {Zhou}},\ }\bibfield  {title} {\bibinfo {title} {Identifying influential nodes in complex networks},\ }\href@noop {} {\bibfield  {journal} {\bibinfo  {journal} {Physica a: Statistical mechanics and its applications}\ }\textbf {\bibinfo {volume} {391}},\ \bibinfo {pages} {1777} (\bibinfo {year} {2012})}\BibitemShut {NoStop}%
\bibitem [{\citenamefont {Spearman}(1987)}]{spearman1987proof}%
  \BibitemOpen
  \bibfield  {author} {\bibinfo {author} {\bibfnamefont {C.}~\bibnamefont {Spearman}},\ }\bibfield  {title} {\bibinfo {title} {The proof and measurement of association between two things},\ }\href@noop {} {\bibfield  {journal} {\bibinfo  {journal} {The American journal of psychology}\ }\textbf {\bibinfo {volume} {100}},\ \bibinfo {pages} {441} (\bibinfo {year} {1987})}\BibitemShut {NoStop}%
\bibitem [{\citenamefont {Buhmann}(2000)}]{buhmann2000radial}%
  \BibitemOpen
  \bibfield  {author} {\bibinfo {author} {\bibfnamefont {M.~D.}\ \bibnamefont {Buhmann}},\ }\bibfield  {title} {\bibinfo {title} {Radial basis functions},\ }\href@noop {} {\bibfield  {journal} {\bibinfo  {journal} {Acta numerica}\ }\textbf {\bibinfo {volume} {9}},\ \bibinfo {pages} {1} (\bibinfo {year} {2000})}\BibitemShut {NoStop}%
\bibitem [{\citenamefont {Sch{\"o}lkopf}\ and\ \citenamefont {Smola}(2002)}]{scholkopf2002learning}%
  \BibitemOpen
  \bibfield  {author} {\bibinfo {author} {\bibfnamefont {B.}~\bibnamefont {Sch{\"o}lkopf}}\ and\ \bibinfo {author} {\bibfnamefont {A.~J.}\ \bibnamefont {Smola}},\ }\href@noop {} {\emph {\bibinfo {title} {Learning with kernels: support vector machines, regularization, optimization, and beyond}}}\ (\bibinfo  {publisher} {MIT press},\ \bibinfo {year} {2002})\BibitemShut {NoStop}%
\bibitem [{\citenamefont {McCulloch}\ and\ \citenamefont {Pitts}(1943)}]{mcculloch1943logical}%
  \BibitemOpen
  \bibfield  {author} {\bibinfo {author} {\bibfnamefont {W.~S.}\ \bibnamefont {McCulloch}}\ and\ \bibinfo {author} {\bibfnamefont {W.}~\bibnamefont {Pitts}},\ }\bibfield  {title} {\bibinfo {title} {A logical calculus of the ideas immanent in nervous activity},\ }\href@noop {} {\bibfield  {journal} {\bibinfo  {journal} {The bulletin of mathematical biophysics}\ }\textbf {\bibinfo {volume} {5}},\ \bibinfo {pages} {115} (\bibinfo {year} {1943})}\BibitemShut {NoStop}%
\bibitem [{\citenamefont {Rosenblatt}(1958)}]{rosenblatt1958perceptron}%
  \BibitemOpen
  \bibfield  {author} {\bibinfo {author} {\bibfnamefont {F.}~\bibnamefont {Rosenblatt}},\ }\bibfield  {title} {\bibinfo {title} {The perceptron: a probabilistic model for information storage and organization in the brain.},\ }\href@noop {} {\bibfield  {journal} {\bibinfo  {journal} {Psychological review}\ }\textbf {\bibinfo {volume} {65}},\ \bibinfo {pages} {386} (\bibinfo {year} {1958})}\BibitemShut {NoStop}%
\bibitem [{\citenamefont {Rumelhart}\ \emph {et~al.}(1986)\citenamefont {Rumelhart}, \citenamefont {Hinton},\ and\ \citenamefont {Williams}}]{rumelhart1986learning}%
  \BibitemOpen
  \bibfield  {author} {\bibinfo {author} {\bibfnamefont {D.~E.}\ \bibnamefont {Rumelhart}}, \bibinfo {author} {\bibfnamefont {G.~E.}\ \bibnamefont {Hinton}},\ and\ \bibinfo {author} {\bibfnamefont {R.~J.}\ \bibnamefont {Williams}},\ }\bibfield  {title} {\bibinfo {title} {Learning representations by back-propagating errors},\ }\href@noop {} {\bibfield  {journal} {\bibinfo  {journal} {nature}\ }\textbf {\bibinfo {volume} {323}},\ \bibinfo {pages} {533} (\bibinfo {year} {1986})}\BibitemShut {NoStop}%
\bibitem [{\citenamefont {Snoek}\ \emph {et~al.}(2012)\citenamefont {Snoek}, \citenamefont {Larochelle},\ and\ \citenamefont {Adams}}]{snoek2012practical}%
  \BibitemOpen
  \bibfield  {author} {\bibinfo {author} {\bibfnamefont {J.}~\bibnamefont {Snoek}}, \bibinfo {author} {\bibfnamefont {H.}~\bibnamefont {Larochelle}},\ and\ \bibinfo {author} {\bibfnamefont {R.~P.}\ \bibnamefont {Adams}},\ }\bibfield  {title} {\bibinfo {title} {Practical bayesian optimization of machine learning algorithms},\ }\href@noop {} {\bibfield  {journal} {\bibinfo  {journal} {Advances in neural information processing systems}\ }\textbf {\bibinfo {volume} {25}} (\bibinfo {year} {2012})}\BibitemShut {NoStop}%
\bibitem [{\citenamefont {Matthews}(1975)}]{matthews1975comparison}%
  \BibitemOpen
  \bibfield  {author} {\bibinfo {author} {\bibfnamefont {B.~W.}\ \bibnamefont {Matthews}},\ }\bibfield  {title} {\bibinfo {title} {Comparison of the predicted and observed secondary structure of t4 phage lysozyme},\ }\href@noop {} {\bibfield  {journal} {\bibinfo  {journal} {Biochimica et Biophysica Acta (BBA)-Protein Structure}\ }\textbf {\bibinfo {volume} {405}},\ \bibinfo {pages} {442} (\bibinfo {year} {1975})}\BibitemShut {NoStop}%
\bibitem [{\citenamefont {Karthika}\ \emph {et~al.}(2016)\citenamefont {Karthika}, \citenamefont {Radhakrishnan},\ and\ \citenamefont {Kalaichelvi}}]{karthika2016review}%
  \BibitemOpen
  \bibfield  {author} {\bibinfo {author} {\bibfnamefont {S.}~\bibnamefont {Karthika}}, \bibinfo {author} {\bibfnamefont {T.}~\bibnamefont {Radhakrishnan}},\ and\ \bibinfo {author} {\bibfnamefont {P.}~\bibnamefont {Kalaichelvi}},\ }\bibfield  {title} {\bibinfo {title} {A review of classical and nonclassical nucleation theories},\ }\href@noop {} {\bibfield  {journal} {\bibinfo  {journal} {Crystal Growth \& Design}\ }\textbf {\bibinfo {volume} {16}},\ \bibinfo {pages} {6663} (\bibinfo {year} {2016})}\BibitemShut {NoStop}%
\bibitem [{\citenamefont {Smeets}\ \emph {et~al.}(2017)\citenamefont {Smeets}, \citenamefont {Finney}, \citenamefont {Habraken}, \citenamefont {Nudelman}, \citenamefont {Friedrich}, \citenamefont {Laven}, \citenamefont {De~Yoreo}, \citenamefont {Rodger},\ and\ \citenamefont {Sommerdijk}}]{smeets2017classical}%
  \BibitemOpen
  \bibfield  {author} {\bibinfo {author} {\bibfnamefont {P.~J.}\ \bibnamefont {Smeets}}, \bibinfo {author} {\bibfnamefont {A.~R.}\ \bibnamefont {Finney}}, \bibinfo {author} {\bibfnamefont {W.~J.}\ \bibnamefont {Habraken}}, \bibinfo {author} {\bibfnamefont {F.}~\bibnamefont {Nudelman}}, \bibinfo {author} {\bibfnamefont {H.}~\bibnamefont {Friedrich}}, \bibinfo {author} {\bibfnamefont {J.}~\bibnamefont {Laven}}, \bibinfo {author} {\bibfnamefont {J.~J.}\ \bibnamefont {De~Yoreo}}, \bibinfo {author} {\bibfnamefont {P.~M.}\ \bibnamefont {Rodger}},\ and\ \bibinfo {author} {\bibfnamefont {N.~A.}\ \bibnamefont {Sommerdijk}},\ }\bibfield  {title} {\bibinfo {title} {A classical view on nonclassical nucleation},\ }\href@noop {} {\bibfield  {journal} {\bibinfo  {journal} {Proceedings of the National Academy of Sciences}\ }\textbf {\bibinfo {volume} {114}},\ \bibinfo {pages} {E7882} (\bibinfo {year} {2017})}\BibitemShut {NoStop}%
\end{thebibliography}
\end{document}